\documentclass[a4paper]{jpconf}
\usepackage{bm,graphicx}
\begin{document}
\title{Neutrinos and electrons in background matter}

\author{Alexander Studenikin}

\address{Department of Theoretical Physics and
Skobeltsyn Institute of Nuclear Physics, Moscow State University,
119992 Moscow, Russia} \ead{studenik@srd.sinp.msu.ru}

\begin{abstract}
We present a rather powerful method in investigations of different
phenomena that can appear when neutrinos and electrons propagate
in background matter. This method is based on the use of the
modified Dirac equations for particles wave functions, in which
the correspondent effective potentials accounting for the matter
influence on particles are included.
\end{abstract}
Within the standard model interaction of electron neutrinos and electrons with
matter composed of neutrons, the modified Dirac equations are
\cite{StuTerPLB05,GriStuTerPLB05,StuJPA06,GriShiStuTerTro12LomCon}
\begin{equation}\label{new_e}
\Big\{ i\gamma_{\mu}\partial^{\mu}-\frac{1}{2}
\gamma_{\mu}(c_l+\gamma_{5}){\widetilde f}^{\mu}-m_l
\Big\}\Psi^{(l)}(x)=0,
\end{equation}
where for the case of neutrinos $m_l=m_\nu$ and $c_l=c_{\nu}=1$, whereas for
electrons $m_l=m_e$ and $c_l=c_e=1-4\sin^{2}\theta_{W}$. For unpolarized matter
$\widetilde{f}^{\mu}=\frac{G_{F}}{\sqrt{2}}(n_n,n_n{\bf v}),$ $n_n$ and
$\mathbf v$ are, respectively, the neutron number density and overage speed. We
have obtained the wave functions for neutrinos and electrons   in the following
form \cite{StuTerPLB05,GriStuTerPLB05,StuJPA06,GriShiStuTerTro12LomCon}
\begin{equation}
\Psi^{(l)}_{\varepsilon, {\bf p},s}({\bf r},t)=\frac{e^{-i(
E^{(l)}_{\varepsilon}t-{\bf p}{\bf r})}}{2L^{\frac{3}{2}}}
\left(%
\begin{array}{c}{\sqrt{1+ \frac{m_l}{ E^{(l)}_
{\varepsilon}-c\alpha_n m_l}}} \ \sqrt{1+s\frac{p_{3}}{p}}
\\
{s \sqrt{1+ \frac{m_l}{ E^{(l)}_{\varepsilon}-c\alpha_n m_l}}} \
\sqrt{1-s\frac{p_{3}}{p}}\ \ e^{i\delta}
\\
{  s\varepsilon\sqrt{1- \frac{m_l}{
E^{(l)}_{\varepsilon}-c\alpha_n m_l}}} \ \sqrt{1+s\frac{p_{3}}{p}}
\\
{\varepsilon\sqrt{1- \frac{m_l}{ E^{(l)}_{\varepsilon}-c\alpha_n
m_l}}} \ \ \sqrt{1-s\frac{p_{3}}{p}}\ e^{i\delta}
\end{array}
\right),
\end{equation}
where the energy spectra are
\begin{equation}\label{Energy_e}
  E_{\varepsilon}^{(l)}=
  \varepsilon \sqrt{{{\bf p}}^{2}\Big(1-s\alpha_n
  \frac{m}{p}\Big)^{2}
  +{m}^2} +c_l {\alpha}_n m_l, \ \ \alpha_n=\pm\frac{1}{2\sqrt{2}}
  {G_F}\frac{n_n}{m_l},
\end{equation}
$p$, $s$ and $\varepsilon$ are particles momenta, helicities and signs of
energy, ``$\pm$" corresponds to $e$ and $\nu_e$.

The developed approach establishes a basis for investigation of different
phenomena which can arise when neutrinos and electrons move in dense media,
including those peculiar for astrophysical and cosmological environments.
Within this approach we have investigated new types of electromagnetic
radiation which can be emitted by neutrinos and electrons moving in dense
matter (the ``spin light" of neutrino and electron in matter, $SL\nu$ and $SLe$
respectively).

\end{document}